# Transfer Loop and Statistical Equilibrium of Korteweg-de Vries-Burgers Systems Associated to Classical Nonlinear Acoustics and Quantum Shock Waves

Jian-Zhou Zhu (朱建州)

*Su-Cheng Centre for Fundamental and Interdisciplinary Sciences, Gaochun, Nanjing 211316, China*

We propose and demonstrate, with the one-dimensional Korteweg-de Vries-Burgers model, the scenarios of transfer loop and *all-scale* statistical equilibrium, the former being associated to shock formation and the latter to Gaussian distributions as in a canonical ensemble, but with wavelength-dependent temperatures. The discussions emphasize, among the multi-disciplinary relevance, the classical nonlinear acoustics and quantum shock waves, for the possibility of more favorable experimental tests.

*Introduction.* — A number of one-dimensional (1D) physical models, including, among others, the Korteweg-de Vries(-Burgers) [KdV(B)], the Navier-Stokes and the quantum fluid equations, present self-advection $u\nabla u$. The addition of the diffusion $\nabla^2 u$, or similarly $-(-\nabla^2)^\ell u$ up to a coefficient $\nu$ ($> 0$), and further the KdV dispersion term $\nabla^3 u$, or similarly $\nabla^{(2m+1)} u$ up to a coefficient $\mu$, constitutes the KdVB equation $\partial_t u + u\nabla u = f - \nu(-\nabla^2)^\ell u + \mu \nabla^{2m+1} u$ with forcing $f$: here we extend the notion to more general integral diffusivity/dissipativity $m$ and dispersivity $\ell$ beyond the normal $m = \ell = 1$. And, among the multi-disciplinary sciences [1], many models associated to the quantum shock wave (QSW) [2–6], have been mapped to KdVB by Kulkarni and Abanov [7] (see also Refs. [8, 9] for quantum and classical plasmas). [Khodel, Kurilkin and Mishutsi [2] actually used precisely the (stationary) KdVB in raising the QSW notion for cold nuclear matter motion.] With the objective of understanding classical turbulence, the KdVB shock has been particularly used to address the intermediate asymptotic issue of inertial spectrum [10].

We consider $u$ on coordinate $x$ with period $L$, normalized to be $2\pi$, and the Fourier coefficient $\hat{u}_k(t) = \mathcal{F}(u) := \int_0^{2\pi} u(x,t) \exp\{\hat{i}kx\} dx/(2\pi)$ with $\hat{i}^2 = -1$. On the 'shell' of wavenumber modulus $n := |k|$ lives the other mode $\hat{u}_{-k}$. The reality of $u$ implies the complex conjugacy (*c.c.*) relation $\hat{u}^*_{-k} = \hat{u}_k$. The objective KdVB system is then

$$(\partial_t + \nu k^{2\ell} - \mu \hat{i} k^{2m+1})\hat{u}_k = \hat{f}_k - \hat{i} \sum_{p+q=k} q\hat{u}_p \hat{u}_q \qquad (1)$$

with $\bar{u} := \hat{u}_0 \equiv 0$ and $f =: \mathcal{F}^{-1}(\hat{f})$ band-limited around some large $n$. Negative sign can be absorbed by $\mu$. Fractional derivatives with real $\ell$ and $m$ [11] can but are not used here.

General $\ell$ and $m$, including the negative integers, may not be completely artificial even in the context of quantum flows: for example, beyond the normal dissipative effect of shock-vortice interplay focused on recently by Mossman et al. [12, 13], the general multi-scale wave-vortex interactions (controlled or not) and possible external interferences can have richer consequences which may be relevant.

Different elements of the last triadic mode-interaction term of Eq. (1) can find appropriate situations to present outstanding physical effects, and one can have the following simple

**Observation 1** *If $f$ is on $n = P$ and $Q$ satisfying $Q - P = 1$, $\hat{u}_{\pm 1}$ can be "indirectly pumped" by $f$ through $\pm Q\hat{i}\hat{u}_{\pm P}\hat{u}_{\mp Q}$ in the last interaction term of Eq. (1).*

Thus we propose that the modulation by the forcing (random or deterministic) in the high-

3wavenumber dissipation range (positive $\ell$) can "indirectly" pump/"beat" on the large scales who then should drive the ("secondary") overall forward transfers to constitute a "transfer loop", forming coherent structures (shocks here). Such transfer loop should be distinguished from the notion of "flux loop" discovered firstly in two-dimensional stratified turbulence [14], because we do not have separate channels, for different energy components, of inverse and forward cascades.
The above transfer loop is closed up by the small-scale damping. However, if the latter is replaced by the large-scale one (negative $\ell$), then small-scale pumping and the quasi-directional "gear" of the self-advection term for "overall" forward transfer lead to "mix-up", resulting in some equilibrium. Somewhat surprisingly, we will see that actually *all* $\hat{u}_k$, including those subject to static forcing or damping, are (nearly) Gaussian!

Consider the (quasi-)1D dynamical problems of some medium, say, QSWs or classical nonlinear acoustics [15, 16] for more specific physical discussions. If strong small-scale modulations are introduced by, say, laser heating, alternating electric field or even some sort of atomic energy release, then they may beat on largest scales and thus the shocks can emerge. The KdVB model is not only physically relevant [2, 7] but also most convenient and simplest for testing our ideas. Our emphasis on quantum fluids is due to the possibly more favorable experiments, like Bendahmane et al.'s [17] remark, related to the light flow experiment on the fiber-based platform, on the "asymmetric case" of the piston Riemann problem.

*The transfer loop.* — The equation for the dynamics of the energy spectrum $E(n) := \sum_{|k|=n} \hat{u}_k^* \hat{u}_k/2 = \hat{u}_k^* \hat{u}_k$, ensemble-averaged ($\langle \bullet \rangle$) when necessary, follows from Eq. (1):

$$\partial_t \langle E(n) \rangle = \langle T(n) \rangle - \nu n^{2\ell} \langle E(n) \rangle + \langle W(n) \rangle. \qquad (2)$$

$T(n) := -\sum_{|k|=n} \hat{u}_k^* \sum_{p+q=k} \hat{u}_p \hat{i} q \hat{u}_q$ is the *transfer (rate) function* and $W(n) := \hat{f}_n \hat{u}_n^* +$ *c.c.* the *work*. We can further define the net flow of energy over $[k_L, k_R]$ by $F(k_L, k_R) := \sum_{k_L \leq n < k_R} T(n)$ and the forward *flux* across $n$ by $\Pi(n) := F(n, \infty)$.

Numerical simulations of various setups of KdVB were carried out with the standard pseudo-spectral method. We first present in Fig. 1 the energy spectra $E(n)$ for $\ell = m = 1$, $f = A[\cos(Px) + \cos(Qx)]$ of constant amplitude $A = 25000$, $\nu = 1/200$ and $Q = P + 1 = K \approx 8192/3$ (associated to the "2/3 rule" for dealiasing), the latter being the largest available wavenumber in the computation (left panel) using $N = 2^{13} = 8192$ grid points. Except for those particularly designated by "KdVB-HMP" and "KdVB-HMP′" to be addressed later, all results are for $\mu = 0$. The left panel presents the spectra at early times $t_1 > t_2$ and in the stationary state ($t_\infty$) from the computation with "high modes pumped" (HMP) by $f$, starting from the double-precision computer noise of a null initial field. The right panel compares the final stationary spectrum from that 8192-simulation with other stationary ones obtained with the same setups except for the resolutions ["32768" for $N = 2^{15}$ grid points, "B-HMP" (representing "Burgers with high-mode pumped with $N = 2^{17}$")], showing that $2^{13}$ grid points are sufficient (to our bare eyes) to ensure the accuracy of the spectrum for wavenumbers before the first turning point at $k_T \approx 1800$. And, $2^{15}$ grid points appear sufficient for the accuracy of the spectrum for all computed wavenumbers, with no visible marked differences to that ("B-HMP") computed from $2^{17}$ grid points.

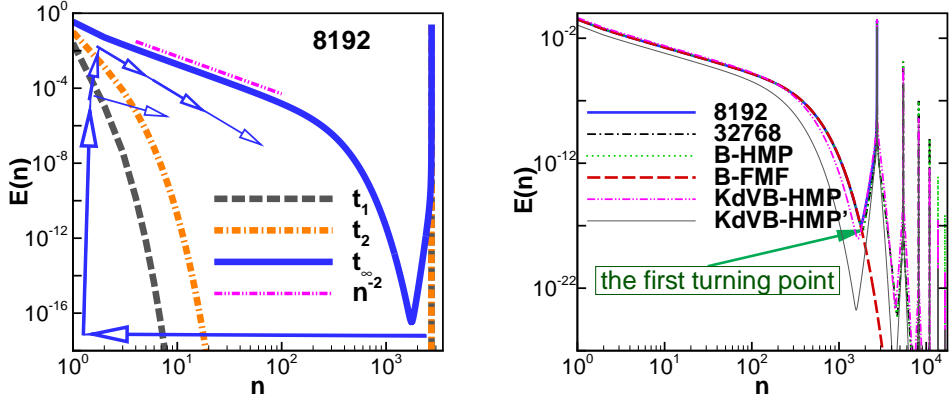

FIG. 1. The velocity power/energy spectra.

The snapshots for the early evolution times $t_1 < t_2$ indicate the transfer-loop scenario, caricatured by the arrows, that the $n = 1$ modes are indirectly but effectively pumped by the forced $\hat{u}_{\pm P}$ and $\hat{u}_{\pm Q}$ and then transfer energy forwardly to larger $n$s, eventually forming an asymptotic inertial scaling $\propto n^{-2}$ corresponding to shock structures. The more conventional $2^{17}$-grid-point simulation, with the first modes fixed ("B-FMF") to be the same $\hat{u}_{\pm 1}$ of the "B-HMP" stationary solution but now without that high-mode pumping, produces no visible differences for $k < k_T$, validating our computations.

The traditional fluxes are given in the upper frame of the figure panel in the left of Table I, with the familiar shape for the conventional "B-FMF" case. The "B-HMP" $\Pi(n)$ however cannot provide the information to explicitly quantify the transfer loop scenario, because the energy transferred (from large $n$s) into lowest-$n$ modes and that transferred out of the latter to larger $n$s cancel each other, resulting in vanishing $\Pi(n)$ for small $n$. We have then plotted for "B-HMP", in the lower frame, the specific mode-mode energy transfer rate functions defined as the particular members of $T(1)$ and $T(n+1)$,

$$R(n-1,n;1) := -\hat{i}[n\hat{u}_{1-n}\hat{u}_n\hat{u}_1^* + (1-n)\hat{u}_n\hat{u}_{1-n}\hat{u}_1^*] + c.c.$$
$$= -\hat{i}\hat{u}_{1-n}\hat{u}_n\hat{u}_1^* + c.c. \quad \forall\, n \geq 1 \tag{3}$$

for the transfer rate of the energy into $\hat{u}_{\pm 1}$ from (the interactions between) $\hat{u}_{\pm n}$ and $\hat{u}_{\pm(n-1)}$, and similarly,

$$R(1,n;n+1) = \begin{cases} -\hat{i}\hat{u}_1\hat{u}_1\hat{u}_2^* + c.c. & n=1 \\ -\hat{i}(n+1)\hat{u}_1\hat{u}_n\hat{u}_{n+1}^* + c.c. & n>1 \end{cases} \tag{4}$$

for the transfer rate of energy into $\hat{u}_{\pm(n+1)}$ from $\hat{u}_{\pm 1}$ and $\hat{u}_{\pm n}$.

TABLE I. $R(n-1,n;1)$ and $R(1,n;n+1)$ graphs (left), and, their figures together with those of $T(n)$ and $\Pi(n)$ (right) at significant values of $n$ for the "B-HMP" case.

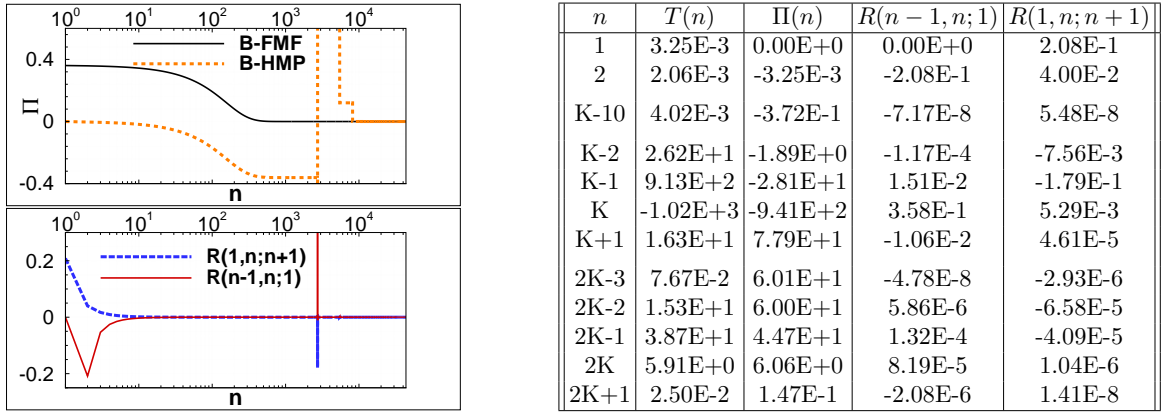

| $n$ | $T(n)$ | $\Pi(n)$ | $R(n-1,n;1)$ | $R(1,n;n+1)$ |
|---|---|---|---|---|
| 1 | 3.25E-3 | 0.00E+0 | 0.00E+0 | 2.08E-1 |
| 2 | 2.06E-3 | -3.25E-3 | -2.08E-1 | 4.00E-2 |
| K-10 | 4.02E-3 | -3.72E-1 | -7.17E-8 | 5.48E-8 |
| K-2 | 2.62E+1 | -1.89E+0 | -1.17E-4 | -7.56E-3 |
| K-1 | 9.13E+2 | -2.81E+1 | 1.51E-2 | -1.79E-1 |
| K | -1.02E+3 | -9.41E+2 | 3.58E-1 | 5.29E-3 |
| K+1 | 1.63E+1 | 7.79E+1 | -1.06E-2 | 4.61E-5 |
| 2K-3 | 7.67E-2 | 6.01E+1 | -4.78E-8 | -2.93E-6 |
| 2K-2 | 1.53E+1 | 6.00E+1 | 5.86E-6 | -6.58E-5 |
| 2K-1 | 3.87E+1 | 4.47E+1 | 1.32E-4 | -4.09E-5 |
| 2K | 5.91E+0 | 6.06E+0 | 8.19E-5 | 1.04E-6 |
| 2K+1 | 2.50E-2 | 1.47E-1 | -2.08E-6 | 1.41E-8 |

For more precise information, Table I tabulates the figures of $T(n)$ and $\Pi(n)$ for $n$ around the pumped wavenumber $K$, and also around $K+K$ where, similarly, modes are indirectly pumped through the triadic interactions: the "cascade" of such indirect pumping leads to the spikes of the spectra. Note that $R(1,1;2) = -R(1,2;1)$ by definition, and we see particularly that $T(1) = 3.25/1000 \ll R(1,1;2) = 2.08/10 < R(K-1,K;1) = 3.58/10$. That is, *the majority of the energy transfer rate function $R(K-1,K;1)$ into $\hat{u}_{\pm 1}$ goes directly through $R(1,1;2)$ into $\hat{u}_{\pm 2}$, while the other minor part non-locally into other higher-$n$ modes*, resulting in the transfer loop.

The above transfer-loop scenario has been checked to present also in KdVB ($\mu \neq 0$ and $m > 0$) and in other cases (not shown) with randomness and/or $f$ on all modes in an interval $[n_1, n_2]$, say, $[K-10, K]$. Note however that if $n_2 - n_1$ is so large that the indirectly pumped interval $[1, n_2 - n_1]$ may be too wide to allow an inertial range for the given setups, the above scenario will be deteriorated unless the resolution is increased sufficiently.

For KdVB with $m > 0$, the oscillations, introduced by the dispersion term, lead to, roughly speaking, more second-order derivative and effectively enhance damping (through the diffusion term) [18], as shown in the right panel of Fig. 1 by the "KdVB-HMP" and "KdVB-HMP$'$" spectra, produced by the same setups as that of "B-HMP" except for the additional term $\mu \partial_x^3 u$ with respectively $\mu = 800000^{-1}$ and $400000^{-1}$. The transfer loop scenario is qualitatively similar to that with $\mu = 0$ and won't be repeated.

*Statistical equilibrium.—* We now replace the standard diffusion by the hypo-diffusion with negative $\ell = -6$ and $\nu = 20^{12}$ for both Burgers and KdVB, the latter with additional hypo-dispersion with $m = -2$ and $\mu = 80^3$, then the large-scale damping will balance the energy injected from small scales, now with $f = \cos(340x) + \cos(339x)$ in our simulation with 4096 grid points for long-time integrations. Fig. 2 presents the time-averaged (over the statistical steady state) $E(n)$ and the probability distribution functions (PDFs) of the imaginary part $s$ of some $\hat{u}_k$s ($s$-PDFs). The PDFs of the real parts are the same, thus not shown, except for a shift of the means for $|k| = 340, 339$ due to our forcing on them.

All the $s$-PDFs appear Gaussian, with their variances indistinguishable for $n$s outside the dissipation range (including those of the pumped modes and beyond), indicating the equipartitioned equilibrium which is consistent with the flat spectrum there [19]. The large-scale dissipation makes the variances decrease with $n$, with no visible deterioration of the Gaussianity [c.f., the lower frame of the upper-right panel for their fluctuations at $k = 6$, about 6th-order smaller than those outside the dissipation range]. Such a result may have some relevance with but is obviously different to the statistical equilibria of the gases of either the KdV solitons [20] or the Galerkin-truncated inviscid Fourier coefficients [21].

Note that the randomness is not directly provided by that of the forcing (which is static), but from the intrinsic chaoticity. And, from Eq. (2), $\langle T(n) \rangle = \nu n^{2\ell} \langle E(n) \rangle \neq 0$ for those unforced modes ($W(n) = 0$) in the (statistical) steady state. However, $|\langle T(n) \rangle|$, now computed by time averaging with the assumption of ergodicity, are several orders smaller than the typical fluctuations of $T$, for both Burgers and KdVB, and have only possible minute influence, if any, on the PDF tails which are noisy from the finite-sample effects. Note also that in the more conventional setups with low-$|k|$ random forcing and high-$|k|$ damping, the corresponding $s$-PDFs at large $|k|$s are far from Gaussian, as presented in the lower-right panel for $s$ at $k = 200$ corresponding to those same setups of "B-HMP" and "KdVB-HMP" in Fig. 1, except for the replacement of the forcing, now at $|k| = 1$ and $2$ with random phases with appropriate amplitudes (the details of which, among others such as the normalizations of the PDFs, are not essential for our discussion).

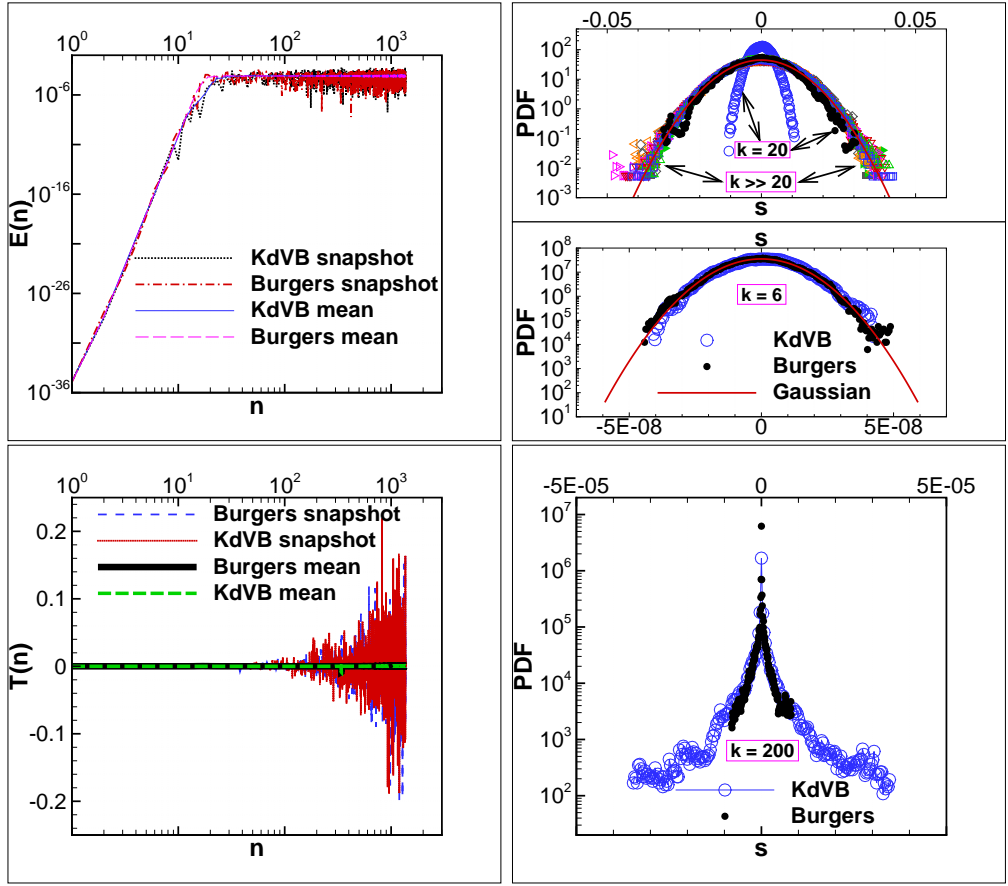

FIG. 2. The mean and snapshot energy spectra (upper-left), spectral transfer functions (lower-left panel) and $s$-PDFs of $\hat{u}_k$ for $k = 50, 80, 210, 340, 341$ and $680$ [various symbols, empty and filled respectively for Burgers and KdVB, well fitted by a single solid line for a Gaussian PDF — the upper frame of the upper-right panel, where also plotted are the $s$-PDFs, again (nearly) Gaussian but clearly different to the others, for $k = 20$ respectively for Burgers and KdVB (filled black circles versus empty blue circles), the former (Burgers) being slightly affected by diffusion to have smaller variance and much more so for the latter (KdVB) subjecting additionally to dispersion.] At $k = 6$ (upper-right panel, lower frame), the $s$-PDFs of Burgers and KdVB are again of nearly the same Gaussian PDFs, in contrast to the typical "non-equilibrium" PDFs (lower-right panel) in the dissipation range of conventional (positive-$\ell$) flows randomly forced at low-$|k|$ modes.

Further remarks follow. First, consistent with previously mentioned indirect pumping on the $n = 1$ modes, $E(1)$ is slightly larger than what would be naively extrapolated from the trend of the spectra (upper-left panel). Second, the hypo-dispersion here is also at work markedly at $n \approx 20$, again enhancing the dissipation; but, no visible differences between the energy spectra of Burgers and KdVB present for ever smaller $n$ ($\lesssim 12$), due to the much stronger (hypo-)dissipation. Third, except for the variances of those of $k$s subject to the marked-dispersion-but-small-dissipation effects (around $|k| = 20$: empty circles versus filled circles in the upper frame of the upper-right panel), the Burgers and KdVB $s$-PDFs are nearly the same, respectively for each $|k|$ with systematic but negligible differences ($\lesssim 2\%$).

Finally, qualitatively the same results apply for cases with the replacement of the deterministic small-scale force by random ones and with other $\ell$ and $m$ with appropriate parameterization (not shown). Different values of $\ell$ and $m$, which together with $\nu$ and $\mu$ constitute the coordinates for the nontrivial "phase diagram" of the dynamics, and different limits, not of the concern of this work, are possible (c.f., e.g., [22]). Our results can be extended in such phase space appropriately, but not arbitrarily. For example, the "weakly damped" KdV case [23] corresponds to $\ell = 0$ and bears the properties somewhat in between our transfer-loop and statistical-equilibrium ones.

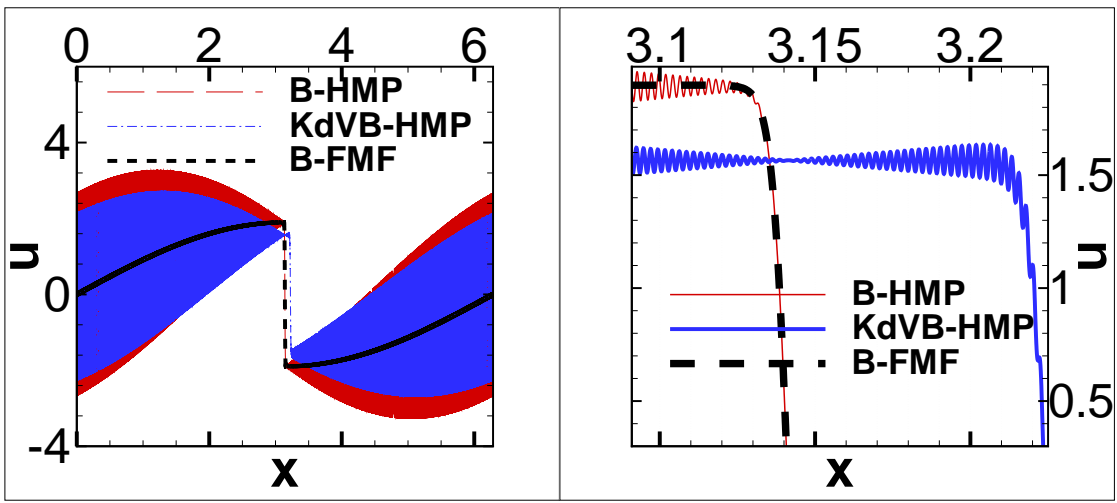

FIG. 3. Stationary profiles of $u$.

*Discussion.—* In terms of Kronecker delta, $\hat{f}_k = \hat{i}A(\delta_{k,K} + \delta_{k,-K} + \delta_{k,K-1} + \delta_{k,1-K})$ from the $f$ described before, so large amplitude $A$ leads, through the interaction term

$\hat{i}\sum_{p+q=k}q\hat{u}_p\hat{u}_q$ in Eq. (1), to the indirect pumping $\Delta\hat{u}$ on $|k|=1$ at least dominantly $\propto \hat{i}$, i.e., $\Delta u = \mathcal{F}^{-1}(\Delta\hat{u}) \propto \sin x$ in configuration space, implying familiarly, as shown in Fig. 3, the canonical B-FMF shock (centered at $x = \pi$) around which are the oscillations (some deails inside the envelop illustrated by the blowup in the right frame) purely introduced by the pumping modulation of B-HMP. The dispersive 'force' $-\hat{i}k^3\hat{u}_k$ of KdVB adds to the indirect pumping on $k = \pm 1$ and leads to more subtle balance with the KdVB-HMP shock location $> \pi$ and with additional oscillations, the latter being seen to vanish at the B-HMP shock center $x = \pi$. The "KdVB-HMP'" and other numerical experiments (not shown) with even larger $\mu$ also present vanishing oscilation amplitude at $\pi$, which is associated to the effect of the 'additional' oscillations as enhancing dissipation (remarked before for the spectra) by reducing the forcing modulation with anti-phase superposition.

The stochastic projected GPE results of Ref. [3] also present strong thermal fluctuations around the mean profile. Such a similarity may be a clue to design alternative approaches [24] to our results in other systems and in more general situations, although the shock and oscillation structures are different in the details for different models and methods.

For $\partial_t \boldsymbol{U} + \boldsymbol{U} \cdot \nabla \boldsymbol{U} = \nu \nabla^2 \boldsymbol{U}$ with $\boldsymbol{U} = \{u, v, w\}$ in $x$-$y$-$z$ space and with $\partial_y u = 0 = \partial_z u$, $u(x)$ solves the self-autonomous 1D Burgers equation, while $v$ and $w$ are respectively affected by $u\partial_x v$ and $u\partial_x w$. Then, the largest scales of $v$ and $w$, even if not both simultaneously forced, can similarly be pumped indirectly by the small-scale forcing. The scenarios proposed then can involve higher-dimensional structures and apply in other similar systems which may be more amenable to experimental tests (c.f., e.g., [13, 15, 17]). Finally, the Observation 1, thus our results, can be extended to systems with interactions of more than three modes (or waves in weak turbulence theory [11]) or higher-order nonlinearities.

## DATA AVAILABILITY

The data that support the findings of this study are available from the corresponding author upon reasonable request.

---